\documentclass[conference]{IEEEtran}
\IEEEoverridecommandlockouts
\usepackage{cite}
\usepackage{amsmath,amssymb,amsfonts}
\usepackage{algorithmic}

\usepackage{graphicx}
\usepackage{subcaption}

\usepackage{textcomp}
\usepackage{xcolor}
\def\BibTeX{{\rm B\kern-.05em{\sc i\kern-.025em b}\kern-.08em
    T\kern-.1667em\lower.7ex\hbox{E}\kern-.125emX}}

\usepackage{url}
\usepackage{hyperref}
\usepackage{comment}

\usepackage{framed}

\usepackage{multirow}
\usepackage{booktabs} % for better lines

\usepackage{caption}
\usepackage{csquotes}

\newcommand{\orcidlink}[1]{%
    \href{https://orcid.org/#1}{%
        \includegraphics[scale=0.35]{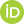}}%
    \hspace{1mm}%
}

\begin{document}
\title{From Obligation to Specification: A Survey on Validating EU AI Act Requirements in RE
% \title{From Regulatory Obligations to Testable Requirements: A Survey on EU AI Act Validation in RE
% \\ {\footnotesize \textsuperscript{*}Note: Sub-titles are not captured in Xplore and
}

\author{
% \IEEEauthorblockN{1\textsuperscript{st} Given Name Surname}
% \and

% \IEEEauthorblockN{\href{https://orcid.org/0000-0002-5396-3543}
%     {\includegraphics[scale=0.35]{icons/orcid_24.png}}\hspace{1mm}T.Y. Emmy Lai}
%     \IEEEauthorblockA{Fraunhofer IAIS\\
%         Sankt Augustin, Germany \\
%         emmy.lai@iais.fraunhofer.de} \\ \\
% \IEEEauthorblockN{\href{https://orcid.org/0000-0002-2691-1406}
%     {\includegraphics[scale=0.35]{icons/orcid_24.png}}\hspace{1mm}Sven Giesselbach}
%     \IEEEauthorblockA{T-System International GmbH\\
%     	Bonn, Germany \\
%     	sven.giesselbach@t-systems.com}
% \and
% \IEEEauthorblockN{\href{https://orcid.org/0000-0002-7640-5237}
%     {\includegraphics[scale=0.35]{icons/orcid_24.png}}\hspace{1mm}Matthias Koch}
%     \IEEEauthorblockA{Fraunhofer IESE \\
%     	Kaiserslautern, Germany \\
%     	matthias.koch@iese.fraunhofer.de} \\ \\

% \IEEEauthorblockN{\href{https://orcid.org/0000-0003-3047-8817}
%     {\includegraphics[scale=0.35]{icons/orcid_24.png}}\hspace{1mm}Héctor Allende-Cid}
%     \IEEEauthorblockA{Fraunhofer IAIS\\
%     	Sankt Augustin, Germany \\
%     	hector.allende-cid@iais.fraunhofer.de} 
    \IEEEauthorblockN{
        \orcidlink{0000-0002-5396-3543}T.Y. Emmy Lai\IEEEauthorrefmark{1},\;
        \orcidlink{0000-0002-2691-1406}Sven Giesselbach\IEEEauthorrefmark{2},\;
        \orcidlink{0000-0002-7640-5237}Matthias Koch\IEEEauthorrefmark{3},\;
        \orcidlink{0000-0003-3047-8817}H\'{e}ctor Allende-Cid\IEEEauthorrefmark{1}
    }
    \IEEEauthorblockA{
        \IEEEauthorrefmark{1}Fraunhofer IAIS, Sankt Augustin, Germany\\
        \IEEEauthorrefmark{2}T-Systems International GmbH, Bonn, Germany\\
        \IEEEauthorrefmark{3}Fraunhofer IESE, Kaiserslautern, Germany\\
        \{emmy.lai, hector.allende-cid\}@iais.fraunhofer.de\;
        sven.giesselbach@t-systems.com\;
        matthias.koch@iese.fraunhofer.de
    }
}

\maketitle

\begin{abstract}   
    With the EU AI Act entering into force, organizations developing or operating AI systems face new obligations on transparency, risk management, and traceability. For Requirements Engineering (RE), these obligations must be translated into testable, auditable requirements and verifiable evidence. However, many organizations currently lack systematic processes to achieve this. We hypothesize that LLM‑based agentic validation tools can support this translation, thereby helping to close this gap.
    
    We present a mixed‑method exploratory study with expert interviews (N=10) and an online survey (N=15) to assess organizational preparedness for EU AI Act‑oriented RE and perceptions of LLM‑based, agentic closed‑loop validation tools, with participants spanning RE, data science, development, and compliance roles. Our results show that, although the EU AI Act is viewed as highly relevant, structured mechanisms to capture regulatory obligations, propagate updates into projects, and maintain lifecycle‑wide traceability and evidence are often missing. Par\-ti\-ci\-pants see LLM‑based tools as promising for mapping obligations to requirements, assessing coverage, and organizing evidence, but express strong concerns about full automation and stress the need for safeguards. Based on these findings, we outline minimum requirements for an EU AI Act‑ready closed‑loop approach.
\end{abstract}


%10 Pages Max + 2 Pages of References
\section{Introduction}
\label{sec:into}
Artificial intelligence (AI) systems are increasingly used in critical domains such as healthcare, public administration, finance, and automotive systems \cite{peraldifrati2010ReInSafetySys,kelly2024navigatingEuAiAct,kilian2025EuAiStandardsChallenges}. For these systems, Requirements Engineering (RE) must address not only functional and traditional quality requirements, but also regulatory, ethical, and risk-related constraints \cite{Kosenkov25SysMapStudyRE,Aberkane21GdprMapStudy,Ayala18OperationalizeGdprRequirements}. The entry into force of the European Artificial Intelligence Act (EU AI Act) defines explicit obligations for “high-risk” AI systems, including requirements for risk management, data governance, documentation, transparency, human oversight, robustness, accuracy, and cybersecurity \cite{eu_ai_act_2023}.

For organizations developing or deploying AI systems in Europe, this introduces a new compliance layer. In terms of RE practice, this creates a moving target: obligations must be reflected in concrete, testable project requirements and evidence artifacts. High-level legal and normative obligations must be translated into high-level requirements (HLRs), refined into testable, low-level requirements (LLRs), and traced to design, implementation, and evidence artifacts throughout the system lifecycle. Because the EU AI Act is recent and standards and guidance are still evolving, many teams must prepare under regulatory uncertainty while ensuring that their RE and quality management processes are 'EU AI Act‑ready'. %At the time of writing, the EU AI Act is not yet fully finalized, detailed guidance and harmonized standards are still emerging. 

Experience from safety-critical and regulated domains suggests that aligning requirements with regulations, maintaining traceability, and validating quality and risk requirements are demanding and often labour-intensive tasks \cite{peraldifrati2010ReInSafetySys}. In the specific context of the EU AI Act, several questions remain open from an RE perspective \cite{kelly2024navigatingEuAiAct, kilian2025EuAiStandardsChallenges, Schnurr2025implementEuAiAct}. It is unclear how far organizations have progressed in building governance structures for AI‑Act‑oriented RE, for example whether they maintain structured views on applicable obligations, how regulatory changes are propagated into RE and quality management processes, and to what extent roles, policies, and monitoring mechanisms have been defined. Likewise, we lack empirical insight into how practitioners validate EU AI Act‑related requirements, which activities they perform, how much effort they spend on mapping obligations and checking coverage, and where they see the greatest risks in the next 12–24 months.

Recent advances in Large Language Models (LLMs) and Multi-Agent Systems (MAS) are being explored to support RE activities \cite{Lai2025PreprintMAS4RE,jin2024mare,sami2024aibasedmultiagentapproach}. These explorations have been complemented by studies such as XTRAREG \cite{Sallam25CaseStudyGdpr} and LLM-based traceability recovery \cite{Hassine24LLMbasedTrace}, which demonstrate the potential of LLMs in extracting and recovering requirements from legal texts and supplementary sources. We envision an LLM-based agentic closed-loop that translates a curated knowledge base on EU AI regulations into structured checklists and validation criteria for RE. Specialized validation agents continuously map HLRs to project-specific LLRs, assess coverage, consistency, and traceability across RE artifacts, and support findings with references to source documents. Human-in-the-loop (HITL) reviews, policy gates, and audit logs feed back into the knowledge base, creating a transparent validation cycle that reduces manual effort while ensuring regulatory control. With this in mind, we hypothesize that an LLM-based agentic closed-loop approach could be used for EU AI Act validation, where agents use structured regulatory knowledge, requirements artifacts, and evaluation data to ensure adherence to the regulation: (i) Map AI Act obligations to project-specific requirements, (ii) Assess coverage, consistency, and traceability, and (iii) Collect and structure evidence in an auditable manner. Conversational agents interact with stakeholders to clarify ambiguities, while HITL mechanisms and policy gates ensure domain expert control and explainable validation results.

The practicality of such tools depends on two interrelated aspects: \emph{readiness} and \emph{acceptance}. In terms of readiness, companies need at least a basic level of governance maturity for AI Act‑oriented RE (e.g., policies, designated responsible persons, monitoring, and initial project experience). In terms of acceptance, practitioners must consider agent‑based, LLM‑based validation useful and trustworthy, understand acceptable error rates and risk classes, and agree on safeguards such as audit logs, source references, HITL verification, controlled deployment, or local processing. Without empirical knowledge on these interrelated aspects, it is difficult to design AI Act‑oriented RE support that fits real‑world constraints and expectations.

In this paper, we report an exploratory survey of practitioners involved in AI‑related RE, data science, engineering, and compliance roles. The survey instrument examines:
    \begin{itemize}
        \item organizational AI‑Act readiness in and around RE (governance, policies, roles, monitoring, timelines)
        \item the current effort and perceived risks mapping and validating regulatory obligations in projects
        \item and attitudes towards LLM‑based, agentic closed-loop validation tools for requirements, including acceptance, trust, perceived usefulness, required guardrails, and willingness to support pilots.
    \end{itemize}
    
Building on this focus, we address the following research questions (RQ):

\textit{\textbf{RQ1:} How prepared are organizations for integrating EU AI Act obligations into RE?}

\textit{\textbf{RQ2:} How do industry practitioners perceive LLM‑based, agentic closed-loop validation tools for requirements in terms of usefulness, trust, and conditions for adoption?}

This paper makes the following contributions:
    \begin{itemize}
        \item \textbf{Empirical snapshot of AI‑Act readiness in RE}. We characterise how organizations currently handle EU AI Act‑related obligations in and around RE, including perceived importance, preparation status, governance structures, and validation practices.
        \item \textbf{Characterisation of current validation effort and challenges}. We provide evidence on how much effort teams invest in mapping and validating regulatory obligations, which EU AI Act requirements they expect to affect RE most, and which risks and challenges they foresee in the next 12–24 months.
        \item \textbf{Industry perspectives on LLM‑based, agentic validation tools}. We investigate how practitioners assess the potential of LLM‑based agents for requirements validation, which RE tasks they would delegate, what error rates and autonomy levels they consider acceptable, and which trust and governance mechanisms (e.g., audit logs, citations, HITL, guardrails) they consider essential.
        \item \textbf{Implications for designing an AI Act‑ready closed-loop for RE}. We derive implications and design considerations for an LLM‑based, agentic closed-loop that supports EU AI Act‑compliant requirements validation in an audit‑ready, ISO 29148‑aligned manner.
    \end{itemize}

The remainder of this paper is structured as follows. Section 2 reviews related work on RE for regulated AI systems, compliance‑oriented RE, and LLM‑based RE support. Section 3 introduces our survey design, instrument, and analysis procedures. Section 4 presents the results on AI Act readiness, current validation practices, and industry perceptions of LLM‑based, agentic validation. Section 5 discusses implications for the design of closed-loop validation tools and outlines limitations. Section 6 concludes and highlights directions for future work.

\section{Related Work}
\label{sec:related-work}
This section situates our work within prior research on regulatory compliance in RE, validation of regulatory requirements, and LLM-based or agentic support for RE.

\subsection{Regulatory Compliance in RE}
\label{sec:rw-regulatory-re}
Regulatory compliance is becoming increasingly important in RE. Several studies have investigated this topic, including a systematic mapping study on automated General Data Protection Regulation (GDPR) compliance in RE using Natural Language Processing (NLP) \cite{Aberkane21GdprMapStudy}. This study found that the intersection of GDPR, NLP, and RE is still largely unexplored, with only one primary study directly combining all three aspects. This indicates that, even for a mature and widely applicable regulation such as the GDPR, RE-related support and empirical understanding of compliance remain limited. Given that the EU AI Act introduces similarly far-reaching obligations for software-intensive systems, we can expect comparable or even more pronounced gaps at the RE level, which our study begins to address.

Another study systematically mapped RE research for regulatory compliance of software-intensive products and services \cite{Kosenkov25SysMapStudyRE}. This study classified RE-related challenges and proposed principles and practices across regulation fields and stakeholders, highlighting missing links between challenges and practices, weak validation, and low industrial involvement. Together with the GDPR-focused work, this suggests that existing research has not yet provided a robust, empirically grounded understanding of how organizations and RE practices cope with concrete regulatory regimes.

More recently, a study analyzed European AI standards and standardization challenges under the EU AI Act based on interviews with 23 organizations \cite{kilian2025EuAiStandardsChallenges}. This study identified tight implementation timelines, high costs, double regulation, and uneven participation in standardization committees as key obstacles, particularly for start-ups and SMEs. While this study underscored the demanding nature of AI Act compliance at the organizational and ecosystem level, it did not address RE-level processes or artifacts explicitly. Our work addresses this gap by empirically characterizing AI Act-related readiness, challenges, and stakeholder perspectives in and around RE, thereby extending prior GDPR- and compliance-oriented research to the specific context of AI regulation.

\subsection{Compliance Validation and Verification for Regulatory Requirements}
\label{sec:rw-compliance-vv}
Several studies have focused on validation and verification for regulatory requirements, particularly for the GDPR. One study reported on constructing a machine-analyzable, model-based representation of the GDPR using UML and OCL to enable automated compliance checking \cite{Torre19CheckingGDPR}. Another study introduced GuideMe, a six-step approach to operationalize GDPR obligations into implementable solution requirements \cite{Ayala18OperationalizeGdprRequirements}. While GuideMe provides more actionable guidance than generic checklists, the authors emphasized limitations in traceability back to legal sources and the absence of continuous, automated assurance beyond requirements quality checks.

Early work is also emerging on compliance validation for the EU AI Act. One study proposed a methodology for interpreting AI Act requirements for high-risk (including safety-critical) AI systems by leveraging product quality models \cite{kelly2024navigatingEuAiAct}. This study demonstrated the approach on an automotive supply-chain use case, mapping legal requirements to quality characteristics and assessment steps. While this offers valuable guidance at the system and product quality level, it does not focus on RE processes or continuous, evidence-based validation workflows. In contrast to these model- and process-based approaches, our work explores practitioners' needs and trust conditions for LLM-based, agentic closed-loop validation of legal obligations such as the EU AI Act at the RE level.

\subsection{Requirements Validation, Traceability, Quality Assurance}
\label{sec:rw-validation-traceability}
A growing body of work addresses requirements validation, traceability, and quality assurance in legal and safety-critical contexts. One study presented a method for expressing non-functional requirements (safety, timing, hardware, performance) and ensuring their validation and traceability in automotive safety-critical systems \cite{peraldifrati2010ReInSafetySys}. Another study compared classification-based and LLM-based prompting approaches for legal requirements traceability, showing that a specialized LLM-based method substantially outperformed generic baselines but still left room for improvement in recall and F2-score \cite{Etezadi25ClasPrompt}. Our study complements this work by examining practitioners' current validation practices and burdens for AI Act-related obligations and deriving minimum requirements for a closed-loop, evidence-based validation approach.

In summary, our study contributes to the existing body of research on regulatory compliance in RE, compliance validation and verification, requirements validation, traceability, and quality assurance, as well as LLM-based and agentic-assisted RE. By exploring practitioners\' needs and trust conditions for LLM-based, agentic closed-loop validation of legal obligations such as the EU AI Act at the RE level, we provide a unique perspective on the challenges and opportunities of using AI and agent-based technologies in regulated settings. Our findings have implications for the development of AI Act-ready closed-loop (AACL), validation tools and highlight the need for further research in this area.

\section{Methodology}
\label{sec:method}
This section describes our mixed-methods research design, including the interview study, survey instrument, and data analysis procedures.

\subsection{Research Method}
With the EU AI Act coming into force and AI systems facing increasing regulation, it becomes important to understand how well organizations are prepared. In particular, \textit{we ask how they integrate the obligations of the EU AI Act into RE}. We also investigate \textit{what practitioners think of LLM-based, agentic tools that support the validation of these obligations}.

%Due to the fact that the topic is still in its early stages and practical applications are only increasingly developing, we chose an exploratory, practice-oriented research design. 
Because the topic is still in its early stages and practical applications are emerging, we chose an exploratory, practice‑oriented research design. We combined two complementary methods:

\begin{itemize}
  \item\textbf{Qualitative Expert Interviews:} to sharpen the research focus, validate the relevance of the topic, and identify central themes and items.
  \item\textbf{Quantitative Structured Online Survey:} to gain a more comprehensive overview of readiness in relation to the AI Act in and around RE and to evaluate points of view toward LLM-based agentic closed-loop validation tools for requirements.
\end{itemize}

\subsection{Study instrument and Participants}
\label{SD}
% Detail the survey instrument (e.g., Likert scales, multiple choice) and sampling method.
\paragraph*{Expert interviews}
We conducted ten interviews with experts from industry and academia. Participants were between 30 and 55 years old and came from two main groups: (i) five practitioners from industry (e.g., requirements engineers, RE instructors, project leads) and (ii) five experts from academia (e.g., professors from software and requirements engineering, legal domain, and AI), primarily based in Germany. Interviews took around 60 minutes and were conducted in a hybrid mode, mostly online. The interviewer took structured notes during and right after each session.

The interviews followed a simple semi-structured outline: (1) Introduction of the participant, their role, and domain. (2) Presentation of the initial research idea on agentic validation of EU AI Act obligations in RE, a closed-loop validation combining foundation models, conversational agents, and HITL. (3) Open discussion and clarification of the idea, including perceived needs, risks, and practical constraints, and finally (4) wrap-up. Insights from these dialogues helped refine the closed‑loop concept, clarify current needs and concerns, and derive candidate topics and items for the survey (e.g., readiness indicators, validation effort, trust mechanisms, autonomy levels, pilot conditions). We applied convenience sampling using our network to acquire the experts.

\paragraph*{Questionnaire survey}
Based on the expert interviews, we designed an anonymous online survey to identify a broader set of perspectives. The survey was implemented in Microsoft Forms and was available for two weeks, from February 29th, 2026, to March 11th, 2026. The survey was mainly conducted in German, with an English version for participants more comfortable in English. %The main language of the questionnaire was German, reflecting the primarily German-speaking target audience, with an English version offered as an option for participants who are more comfortable in English. 

On the first page of the survey, participants received an information and consent statement covering study purpose, duration (about 20-30 minutes), data protection, storage periods, and their rights. Only participants who agreed could proceed. The survey\footnote{The full survey instrument and the corresponding codebook will be published in an online replication package upon acceptance of this paper.} comprises the following main components:
\begin{itemize}
  \item \textbf{Profile:} role(s), sector, experience with AI/data-science projects, position in the AI system lifecycle, high-risk context (EU AI Act), and organization size.
  \item \textbf{General handling of legal and normative requirements:} governance, processes for propagating regulatory changes into RE/QMS, and auditability.
  \item \textbf{EU--AI--Act status and preparation:} personal knowledge, perceived importance, internal activities and timelines, policies and owners, expected applicability, and monitoring of emerging norms and guidance.
  \item \textbf{Data and validation context:} perceived data readiness, main barriers to data use, evaluation dataset lead times, annotation practices, strategies when data are insufficient.
  \item \textbf{Compliance impact and effort:} current and expected effort for compliance mapping and validation, recent audit findings, obligations with strongest impact on RE, and perceived risks of non-compliance.
  \item \textbf{AI, agents, and HITL:} acceptance of AI/agent-based support in RE, perceived usefulness and user-friendliness, trust and risk tolerance, tolerated error rates and autonomy levels, and prioritized trust mechanisms (e.g., audit trails, citations, guardrails, HITL).
  \item \textbf{Closed-loop validation concept:} perception of current validation gaps, expected benefits of an agent-based closed-loop, necessary technical and organizational conditions, success factors for a pilot project, and perceived challenges.
\end{itemize}

We mainly used structured question formats. These included 5‑point Likert scales \cite{joshi2015likert} (1 = \enquote{Strongly disagree}, 5 = \enquote{Strongly agree}) with an additional \enquote{Don't know/Not applicable} option, single‑choice questions for categorical attributes such as company size or self‑assessed knowledge level, and multiple‑choice questions with an upper limit (e.g., \enquote{Select up to 3 options}) for practices, barriers, and priorities. We also added a small number of open‑ended questions, particularly on the biggest challenges, missing evidence, and perceived improvements.

The link to the survey was shared through various channels, including social media, specialized mailing lists, and direct outreach to experts in the field, in an effort to reach a diverse group of respondents and minimize bias in the results. The target group was experts who deal with RE and related areas, such as business analysts, AI/ML engineers, data scientists, compliance/legal departments, quality/QMS, and project management, for AI development projects.

\subsection{Data Analysis Plan}
% Explain how the data will be processed (e.g., using software like Excel, SPSS, R).
We analyzed the data in two complementary streams: qualitative analysis of open-ended answers and interview notes using in-vivo coding, and quantitative analysis of the structured survey responses.

\paragraph{Qualitative analysis of open-ended responses} Free-text answers were anonymized and analyzed using an in-vivo coding approach. In a first pass, the author extracted characteristic phrases from the respondents' wording as in-vivo codes. In later passes, similar codes were grouped into higher-level themes, such as perceived AI‑Act‑related pain points in RE and validation, missing evidence and data, expectations towards LLM-based, agentic closed-loop tools, and perceived barriers and prerequisites for pilot projects. These qualitative themes are used to contextualize the quantitative findings, to illustrate typical concerns and expectations, and to derive implications for the design of an AI‑Act‑ready closed-loop validation approach.

\paragraph{Quantitative analysis of structured survey response} 
Responses without a declaration of consent and incomplete submissions are not included in the analysis, as they are not saved in advance. Answer options such as "Don't know/Not applicable" (\emph{Weiß nicht/Nicht anwendbar}), or "Unclear/No information provided" (\emph{Unklar/Keine Angabe}) were treated as separate categories in frequency tables and visualizations, but excluded from the computation of numerical summaries (e.g., medians) and from aggregated indicators. Considering the small sample (N = 15) and exploratory study design, we focus on descriptive analyses rather than hypothesis testing. %Considering the (small) sample size (N = 15) and the exploratory nature of the study, our analysis focuses on more descriptive and exploratory steps than on confirming hypothesis tests.

\paragraph{Data preparation} We mapped Likert-scale items to numeric values from 1 to 5 and grouped related items into thematic blocks, such as AI‑Act‑related governance and readiness, current compliance effort and perceived impact, and attitudes towards AI- and agent-based closed-loop tools.

\paragraph{Descriptive statistics and visualization} For all items and composite indicators, we calculated descriptive statistics: absolute and relative frequencies for categorical and multiple-choice items, and medians and interquartile ranges for Likert-scale items. We created charts to visualize the distributions.

\paragraph{Exploratory subgroup comparisons} To explore possible differences between contexts, we performed exploratory subgroup (e.g.\ RE versus other roles, high‑risk versus non‑high‑risk) analyses based on selected attributes. For these subgroups, we compared distributions of key items and three composite 0–100 indices (AI‑Act readiness, compliance burden, and AI/agent acceptance), which aggregate normalized Likert and ordinal items, treating “Don’t know/Not applicable” responses (coded as 0) as a separate category in descriptive statistics. For each subgroup, we inspected cross‑tabulations and simple distribution summaries (medians, interquartile ranges, and boxplots) to identify qualitative patterns in these indices and in selected single items. Because of the small sample size and the use of rating scales, we did not conduct formal significance tests; all subgroup findings are interpreted qualitatively and as exploratory patterns.
%To explore possible differences between contexts, we performed exploratory subgroup (e.g. RE against other roles) analyses based on selected attributes. For these subgroups, we compared distributions of key items and composite indicators using cross-tabulations. Because of the small sample size and the use of rating scales, we did not conduct formal significance tests. All subgroup findings are interpreted qualitatively and as exploratory patterns.\\

\subsection{Threats to Validity}
While we took steps to mitigate common threats to validity in our mixed-methods empirical study \cite{creswell2017research}, there are still important limitations to consider when interpreting our findings.

\paragraph{Internal Validity}
Internal validity refers to how well our results reflect the underlying situation, rather than being influenced by errors or biases. Our study relies on self‑reported interview and survey data, which may be affected by memory and social desirability bias. We mitigated this by ensuring anonymity, using neutral wording, and allowing “Don’t know/Not applicable” responses, but some over‑optimistic assessments of readiness cannot be ruled out. In addition, our use of structured interview notes instead of verbatim transcripts limits the granularity and nuance we can analyze.

%Our study relies on self-reported information from expert interviews and surveys, which can be subject to memory bias and social desirability bias. To minimize these effects, we ensured anonymity, used neutral questions, and provided options for participants to indicate uncertainty or lack of knowledge. However, we cannot rule out the possibility that responses may be biased towards more positive or normative views of organizational readiness. Additionally, our reliance on structured notes from interviews may have limited our ability to capture nuance and exact wording.

\paragraph{External Validity}
External validity refers to how generalizable our results are beyond our sample. We used convenience sampling within existing expert networks and obtained a small, self‑selected sample, biased towards organizations already interested in the EU AI Act and AI‑based tools. Our results should therefore be interpreted as an initial snapshot of this group rather than representative for all organizations, sectors, or countries. We address this by clearly describing our sample and framing our conclusions as exploratory and hypothesis‑generating.
%Our study used convenience sampling through expert networks and targeted contacts, resulting in a limited sample size with a focus on initial users interested in the EU AI Act and AI-based tools. Therefore, our results should be considered a preliminary overview of this specific group and not representative of all organizations, industries, or countries. We address this limitation by clearly describing our sample and presenting our conclusions as preliminary and useful for generating further questions and hypotheses.

\paragraph{Construct Validity}
Construct validity refers to how well our questions and coding measure the concepts we care about, such as AI Act readiness and acceptance of closed-loop approaches. We supported construct validity by defining key terms, grouping related questions into topical blocks, and refining items based on expert feedback from the qualitative pre‑study. However, we did not conduct formal scale validation, some complex constructs are covered by few items, and qualitative coding was performed without independent double‑coding. As a result, composite indicators should be seen as heuristic summaries rather than validated measurement scales.
%To support construct validity, we defined key terms and grouped related questions into topical blocks. We also refined items and answer options based on expert feedback from a qualitative pre-study. However, we did not conduct formal item validation, and some complex concepts are covered by only a few questions. Therefore, our aggregate indicators should be considered heuristic summaries rather than validated scales. Additionally, our qualitative analysis was performed by two authors without independent double-coding, which increases the risk of researcher bias.

\paragraph{Conclusion Validity}
Conclusion validity refers to how reasonable our conclusions about relationships and differences are. Given the small sample size and the use of rating scales, we restrict our analysis to descriptive statistics and exploratory subgroup comparisons. We deliberately avoid formal hypothesis testing and causal claims and instead interpret observed patterns as tentative indications. Larger and more diverse samples will be needed to confirm or refine the relationships suggested by our results. %Due to our small sample size and use of rating scales, we can only identify and quantify relationships between variables to a limited extent. Therefore, we focus on descriptive statistics, simple composite indicators, and exploratory comparisons between subgroups. We avoid formal hypothesis testing and strong causal claims, and instead interpret differences between subgroups as qualitative patterns suggesting possible relationships. Future work will require larger and more diverse samples to replicate and extend our findings.

Despite these limitations, we believe that our study provides useful initial insights into the acceptance of AI laws in the field of RE and the current views of experts towards LLM-based agentic closed-loop validation tools.

\section{Results and Key Findings}
\label{sec:results}
This section reports our empirical findings on AI Act readiness, compliance effort and risks, and practitioners’ acceptance of LLM-based, agentic validation tools.

\subsection{Characterizing the Survey Respondents}
\begin{figure*}[h]
        \centering
        \includegraphics[width=\textwidth]{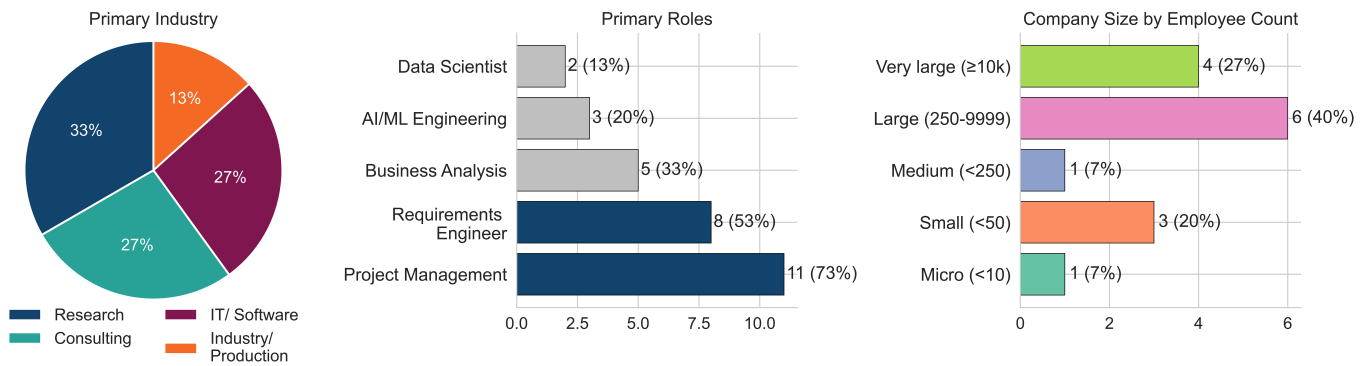}
        \caption{\textit{Cohort overview:} Participants’ backgrounds, showing the distribution of sectors (single choice), primary roles (multiple choice), and company sizes (single choice) in the study cohort (N=15).}\vspace{-10pt}
    \label{fig:cohort-overview}
    \end{figure*}

The survey includes 15 professionals working in IT‑related fields. The respondents come from a variety of sectors, including research (33\%), consulting (27\%), IT and software (20\%), industry and manufacturing (13\%), and IT for industry and manufacturing (7\%). Most of the organizations are large enterprises, with 40\% having 250-9{,}999 employees, 27\% having over 10{,}000 employees, and the remainder being small or medium‑sized enterprises (including micro‑enterprises; cf. Figure \ref{fig:cohort-overview}).
    
The respondents have varying levels of experience with AI and data science projects, with 40\% having 2-3 years of experience, 40\% having more than 6 years of experience, and the remainder having less experience. Many of the respondents (73\%) are currently working on projects that are affected by the EU AI Act, and 33\% are working in high-risk contexts.

The respondents hold a range of roles, including project management, RE, business analysis, and ML/AI engineering. They often work as consultants, internal users, developers, providers, and deployers/operators. This suggests that the respondents are involved in various aspects of the AI system life cycle, including specification, delivery, and advisory functions. 

\subsection{Survey Findings}
To conduct a targeted evaluation of the survey, we divided the 58 questions into three main categories:
(i) organizational readiness for the EU AI Act, 
(ii) compliance effort, data, and risk perception, 
and (iii) acceptance of AI agent-based tooling in RE.
Categories (i) and (ii) primarily address RQ1 (organizational preparedness), while category (iii) addresses RQ2 (perceptions of LLM‑based, agentic closed-loop validation tools).

\subsubsection{EU AI Act Readiness}
\begin{figure}[h]
        \centering
        \includegraphics[width=0.5\textwidth]{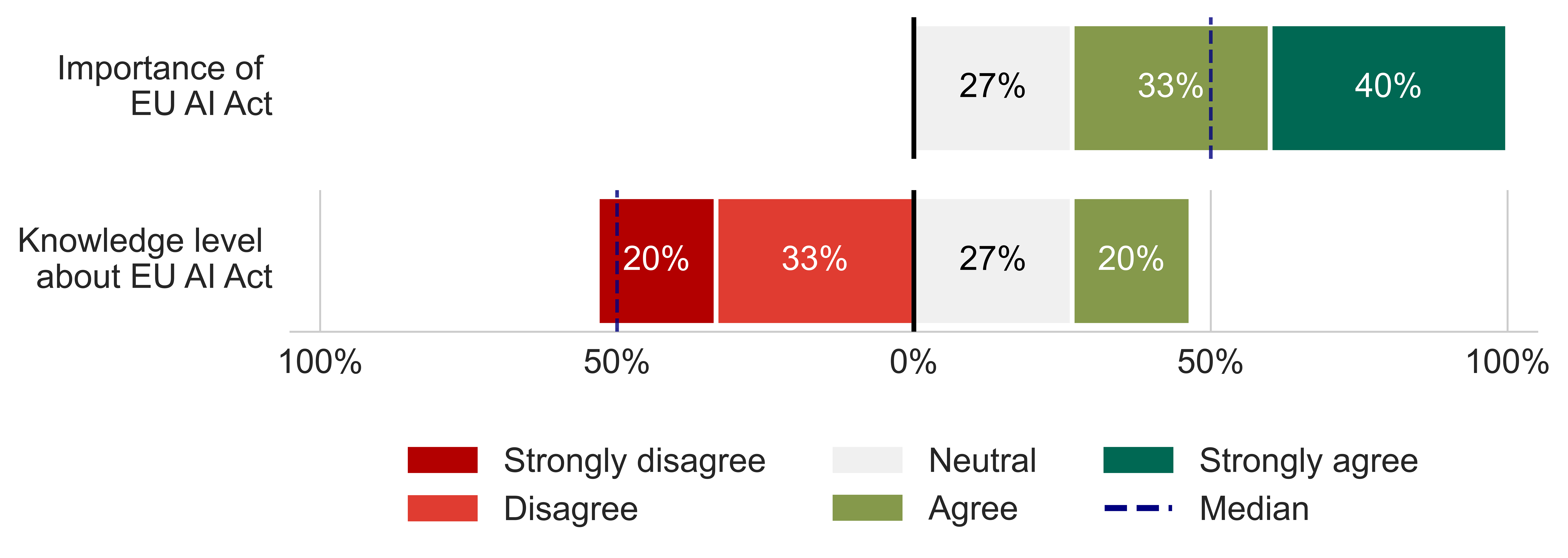}
        \caption{\textit{Knowledge and Importance of the EU AI Act:} Survey responses (N=15) comparing participants’ self‑reported knowledge level about the EU AI Act with their perceived importance of the regulation, shown as stacked Likert scale percentages with indicated medians.}\vspace{-12pt}
        \label{fig:ai-act-knowledge}
    \end{figure}

Addressing RQ1, the survey shows that the EU AI Act is considered highly relevant by the participants. Almost all respondents (93\%) rate it as at least "rather important" for their organizations or projects, and a significant share (47\%) classify it as "very important". However, self-reported knowledge of the Act is moderate, with most participants describing their level as "basic knowledge" (33\%). The remaining responses are split between clearly low and clearly high self-assessed knowledge, so that roughly two-fifths feel confident about the regulation while about one-third still see themselves at a low knowledge level (cf. Figure~\ref{fig:ai-act-knowledge}). Figure~\ref{fig:ai-act-knowledge} summarizes this pattern by juxtaposing the stacked Likert ratings for importance and knowledge, including their medians.
    
\begin{figure*}[h]
        \centering %, height=0.5\textheight
        \includegraphics[width=0.8\textwidth]{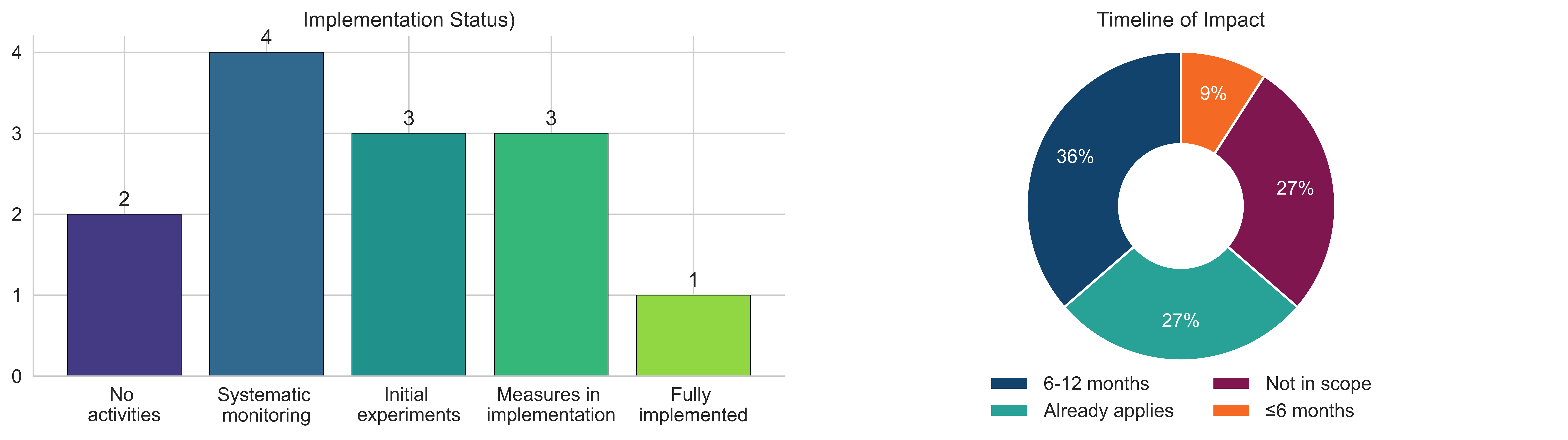}
        \caption{\textit{AI‑Act Readiness Overview:} AI‑Act readiness overview showing the current implementation status of required measures and the expected timeline of regulatory impact across surveyed projects.}\vspace{-10pt}
        \label{fig:ai-act-readiness-overview}
    \end{figure*}

%On the organizational level, companies are mostly in the early to intermediate stages of responding to the Act. A majority (27\%) systematically monitor AI Act developments and emerging guidelines, and six organizations (40\%) are either running first implementation experiments or actively adapting processes and artifacts. Only one organization (7\%) reports that internal requirements are already fully implemented, embedded, and monitored on a continuous improvement basis. 
Organizationally, most companies are in early to intermediate stages of responding to the Act: 27\% systematically monitor developments, 40\% are experimenting with implementation or adapting processes, and only one organization reports fully implemented and monitored internal requirements. At the same time, a non‑negligible subset either has not yet started concrete preparation activities or is unsure about their status, and expectations about the onset of obligations range from ``already applicable'' to predominantly within the next one to three years, with some projects currently judged to remain out of scope. This distribution of implementation status and expected timeline of regulatory impact across the surveyed projects is depicted in Figure~\ref{fig:ai-act-readiness-overview}.
    
In current projects, compliance handling remains largely manual. The dominant practices are documented checklist/table/wiki-based reviews (40\%) and the use of AI in controlled environments with prompts, policies, and audit trails (40\%). Automated evidence collection or CI/CD-integrated checks are not yet in use. Regulations related to EU AI legislation are often still in the development process, with several organizations reporting policies in draft or interim status rather than fully enforced and audited.

% \begin{framed}
% \paragraph*{Key finding}
% The EU AI Act is considered highly important, but most organizations are still in the early stages of preparation. They are monitoring developments and experimenting with implementation, but compliance handling remains largely manual, and only a few have fully implemented and continuously monitored policies.
% \end{framed}
\begin{framed}
\paragraph*{Key finding}
The EU AI Act is viewed as highly important, but most organizations are still in early preparation stages; they mainly rely on manual, checklist‑based practices, and only few have fully implemented and continuously monitored policies.
\end{framed}

\subsubsection{Compliance Effort, Data, and Risks}
Also addressing RQ1, the time spent on mapping and validating regulatory obligations, such as the EU AI Act and GDPR, is currently relatively low for most respondents (47\% report less than 10 hours per month). However, those who provided estimates expect this time to increase significantly as the Act becomes fully applicable, with some anticipating a 20-49\% or even 50\% increase.
  
\begin{figure*}[h]
        \centering
        \includegraphics[width=\textwidth]{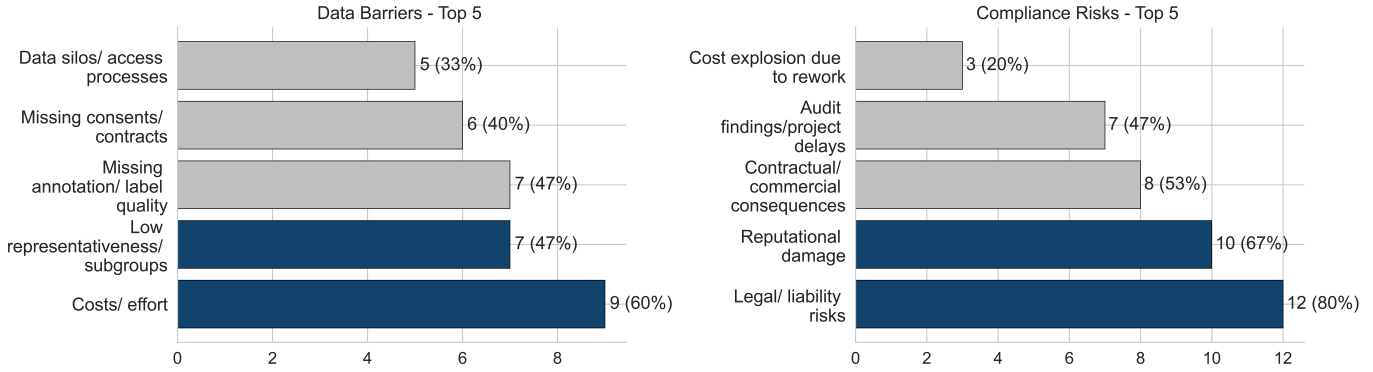}
        \caption{\textit{Barriers and Risks:} top perceived data barriers and compliance risks for adopting the proposed validation approach  (multiple choice), showing the five most frequently selected items by number and percentage of participants.}\vspace{-10pt}
        \label{fig:risks-data-barriers}
    \end{figure*}

The main concerns about non-compliance are legal and liability risks (80\%), reputational harm (67\%), and contractual or commercial consequences (53\%). Operational disruptions and data/privacy incidents are less frequently mentioned. In addition, some participants anticipate more indirect compliance risks, such as negative audit findings, project delays, or substantial rework costs if issues are discovered late (cf. Figure~\ref{fig:risks-data-barriers}).
    
When it comes to data readiness, participants identify several key obstacles, including:
    \begin{itemize}
        \item Cost and effort (9 responses)
        \item Insufficient annotation/label quality and sub-group coverage (7 responses each)
        \item Organizational/data access barriers, such as data silos and processes (5 responses)
    \end{itemize}
    
Figure~\ref{fig:risks-data-barriers} further shows that limited availability of consent or contractual coverage for using data joins these issues among the most frequently cited data barriers, with cost and effort clearly standing out as the dominant concern. 

Qualitative responses highlight that accessing realistic customer data and well-governed evaluation sets are often a major bottleneck. These findings suggest that organizations expect to spend more time on regulatory compliance in the future, but currently lack robust and well-curated evidence bases for systematic validation.

% \begin{framed}
% \paragraph*{Key finding}
% While the current time spent on regulatory compliance is relatively low, respondents expect a significant increase in effort once the AI Act becomes fully applicable. Meanwhile, practitioners are concerned about legal and reputational risks and face challenges in accessing high-quality, representative, and well-governed data and evidence.
% \end{framed}
\begin{framed}
\paragraph*{Key finding}
Current efforts for mapping and validating regulatory obligations is still low, yet respondents expect a substantial increase once the AI Act becomes fully applicable, while simultaneously facing significant legal and reputational risks and persistent data and evidence gaps.
\end{framed}

\subsubsection{AI Agent-Based Acceptance}
Addressing RQ2, the majority of respondents (93\%) consider the topic of AI, agents, and HITL to be highly relevant. They expect AI/agent support in RE to bring significant gains in quality and/or efficiency, and 11 of 15 participants plan to adopt such tools within the next 6–12 months. 

Respondents are most willing to delegate tasks such as drafting and structuring requirements (80\%), generating test cases and acceptance criteria (67\%), and documentation/traceability (60\%) to AI agents. They also see potential in using AI agents to map EU AI Act obligations to RE artifacts (40\%).

However, acceptance of AI-based agents is conditional on strong safeguards. Respondents emphasize the need for low error rates, which should not exceed 10\% in most cases. Autonomous decisions without human approval are generally only accepted in low-risk use cases (53\%). The preferred operational mode is a co-pilot with mandatory human approval (53\%), followed by suggestion-only modes (20\%).

\begin{figure*}[h]
        \centering
        \includegraphics[width=\textwidth]{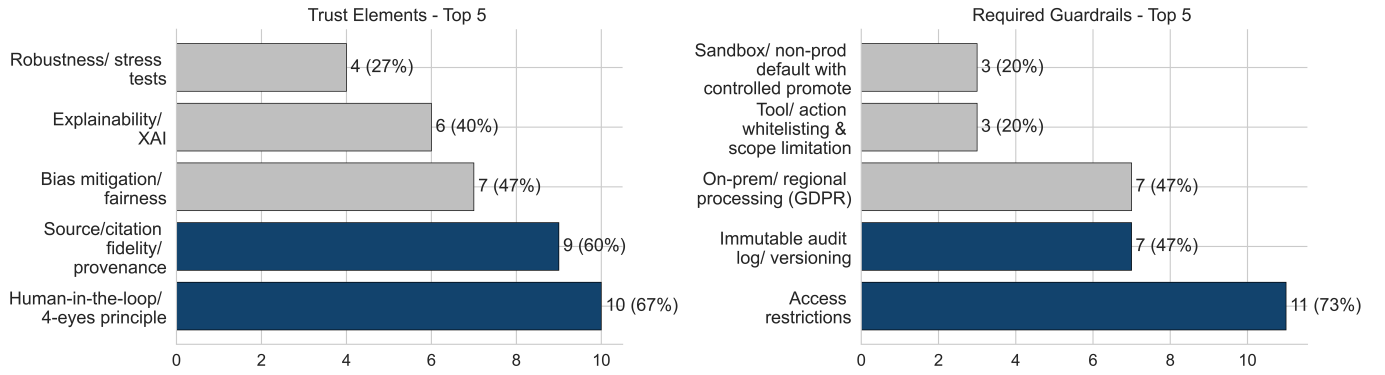}
        \caption{\textit{Key Requirements and Guardrails:} Top‑ranked trust elements and required guardrails for the proposed validation approach (number and share of participants selecting each item). (Left) Trust Elements (Top 5): Human oversight, provenance, bias mitigation, explainability, and robustness. (Right) Required Guardrails (Top 5): Access control, audit logging, internal processing, scope limitation, and sandboxing.}\vspace{-10pt}
        \label{fig:ai-agent-guardrails}
    \end{figure*}

Trust in AI agents hinges on transparency and controllability. Respondents consider faithful source citations (60\%), HITL principles (67\%), explainability, and bias mitigation to be essential trust enablers. They also demand strict guardrails, including least-privilege access control (73\% of selections), immutable audit logs (47\%), and on-premise or regionally constrained processing to meet GDPR expectations (47\%). In addition, respondents highlight robustness and stress testing as further trust elements and call for technical containment measures such as tool isolation with explicit scope limitations and, for some, sandboxed or tightly controlled prompts as additional guardrails.  These top-ranked trust requirements and guardrails are summarized in Figure~\ref{fig:ai-agent-guardrails}.
    
\begin{figure}[h]
    \centering
    \begin{subfigure}[b]{0.5\textwidth}
        \includegraphics[width=\textwidth]{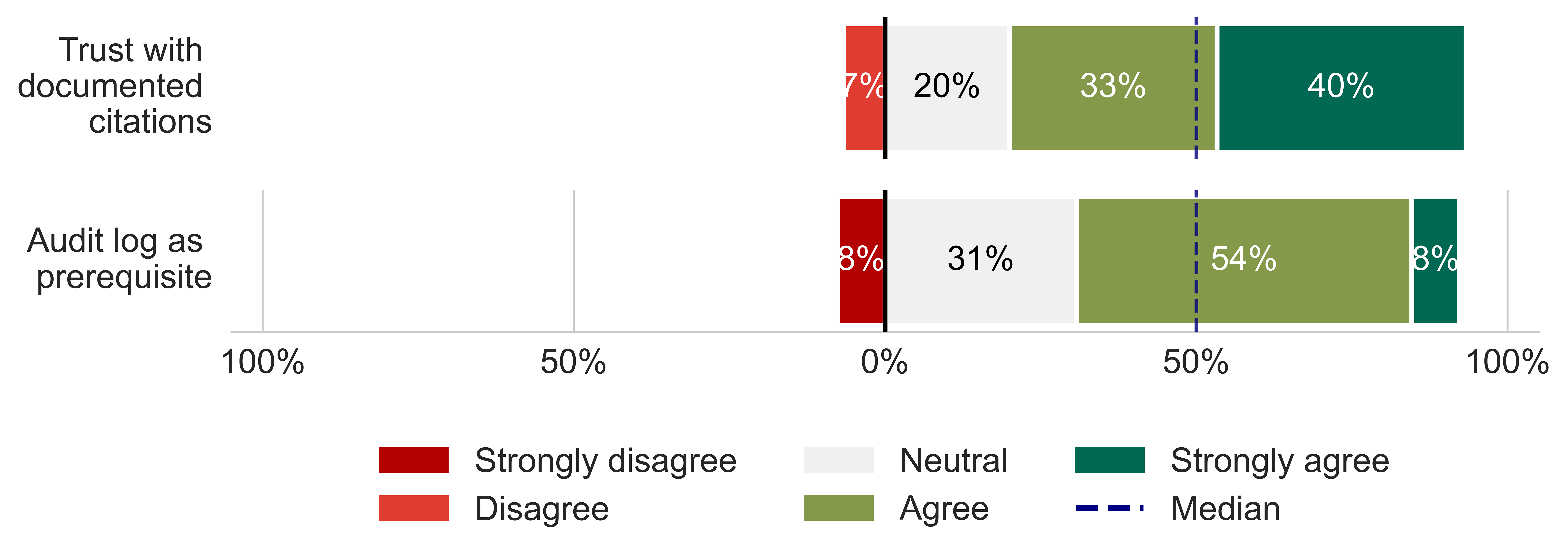}
        \caption{Trust \& Transparency Requirements: Participants’ agreement that documented citations and an audit log are prerequisites for trusting the approach.}
        \label{fig:TrustTransparent}
    \end{subfigure}
    \begin{subfigure}[b]{0.5\textwidth}
        \includegraphics[width=\textwidth]{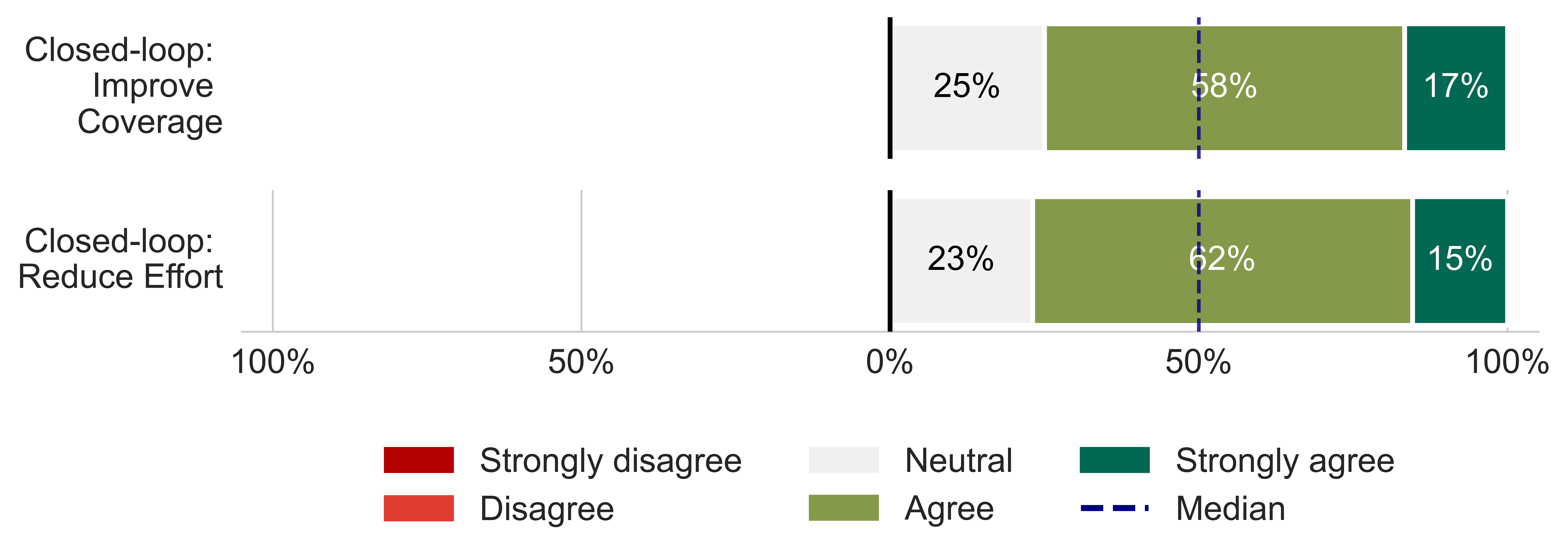}
        \caption{Closed-Loop Potential \& Willingness: Participants’ perceptions of whether a closed-loop setup can improve coverage and reduce effort.}
        \label{fig:PotentialWillingness}
    \end{subfigure}
    \begin{subfigure}[b]{0.5\textwidth}
        \includegraphics[width=\textwidth]{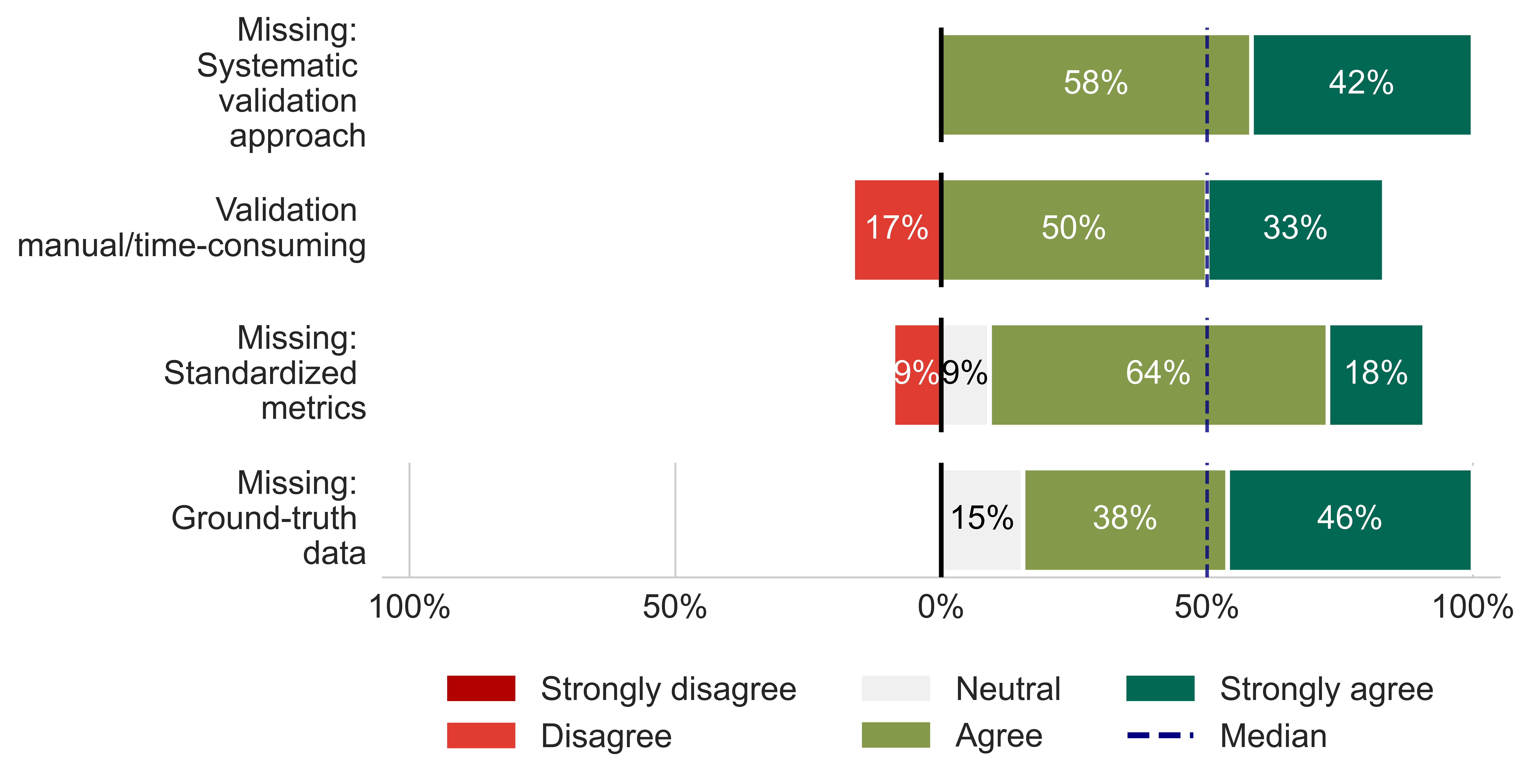}
        \caption{Current Validation Problems: Participants’ agreement with statements on missing systematic validation, time‑consuming manual validation, missing standardized metrics, and missing ground‑truth data.}
        \label{fig:ValidationProblem}
    \end{subfigure}
    \caption{\textit{Perceptions of the Proposed Validation Approach:} Survey responses on trust and transparency requirements, current validation problems, and perceived closed-loop potential and willingness for the proposed validation approach (stacked Likert scales with percentage distribution and median).}\vspace{-12pt}
    \label{fig:acceptance}
\end{figure}

Many respondents (53\%) express willingness to support pilots for agent-based closed-loop approaches, with expectations of at least a 10-30\% reduction in time to compliance baseline and/or improved coverage and quality of mappings. The corresponding Likert-scale responses indicate that most participants agree that documented citations and an audit log are prerequisites for trusting such an approach and that a closed-loop setup can both improve coverage and reduce effort, while simultaneously revealing widespread agreement that current validation is largely manual, lacks a systematic approach, standardized metrics, and sufficient ground-truth data (cf. Figures~\ref{fig:TrustTransparent}–\ref{fig:ValidationProblem}) Figure~\ref{fig:acceptance} provides an overview of these perceptions, combining trust and transparency requirements, current validation problems, and perceived closed-loop potential and willingness.

% \begin{framed}
% \paragraph*{Key finding}
% Practitioners show a high interest in adopting AI/agent-based support in RE, particularly for tasks like drafting, test case generation, and documentation. However, they require a co-pilot mode with human approval, low error rates, and strong transparency and security measures to ensure trust and acceptance.
% \end{framed}
\begin{framed}
\paragraph*{Key finding}
Practitioners are highly interested in AI‑ and agent‑based support for RE tasks such as drafting, test case generation, and documentation, but only under a co‑pilot mode with human approval, low error rates, and strong transparency, security, and governance safeguards.
\end{framed}

%%%%%%%%%% QUALITATIVE ANALYSIS %%%%%%%%%%%%
\subsubsection{Qualitative Analysis}
%The open-ended survey responses and expert interviews provide additional context to the quantitative findings. The detailed coding procedure is described in Section~\ref{SD}. 
We identified four recurring themes:

\paragraph{Delivery Pressure and Limited Capacity}
Several survey respondents describe a tension between acknowledging the importance of the EU AI Act and lacking the time and resources to address it systematically. One participant summarizes their main challenge as having the capacity to implement AI Act obligations while meeting daily delivery pressures. This is reflected by industry interviewees, who report that AI Act-related work is often handled as a best-effort activity rather than a structured, continuous process.

\paragraph{Missing Evidence and Access to Realistic Data}
A second theme concerns the lack of well-governed evidence and access to realistic data, particularly when it comes to real-world customer data. One respondent notes that obtaining customer consent is a significant hurdle. Interviewees also emphasize that while organizations may have large amounts of operational data, consent, contractual restrictions, and data protection concerns make it difficult to reuse them as curated, versioned evidence for validation.

\paragraph{Uncertainty in Interpreting AI Act Obligations}
Respondents highlight uncertainty about how to interpret AI Act provisions in concrete projects and a lack of systematic, testable validation approaches. One participant identifies the need for automatic identification of relevant law changes, while another stresses the need for extraction of context-relevant legal requirements to ensure that nothing essential is overlooked. Interviewees also point to long-standing weaknesses in validation and quality assurance, and call for more systematic checks of quality requirements.

\paragraph{Expectations for Agent-Based Closed-Loop Support}
Finally, both surveys and interviews reveal clear expectations and boundary conditions for agent-based, LLM-supported tooling. Respondents express a desire for tools that can automatically identify relevant law changes and translate legal requirements into context-specific checks, but only under strict guardrails such as on-premise deployment, audit trails, and human review gates. Interviewees describe the closed-loop idea as promising, but stress the need to focus on a concrete, tractable problem and involve domain experts in the HITL validation.

% \begin{framed}
% \paragraph*{Key finding}
% The qualitative findings converge on four main concerns: delivery pressure and limited capacity, missing evidence and access to realistic data, uncertainty in interpreting AI Act obligations, and a perceived gap in systematic, tool-supported mapping and validation approaches. At the same time, there is a clear interest in agent-based closed-loop support under strict HITL and transparency constraints.
% \end{framed}
\begin{framed}
\paragraph*{Key finding}
Qualitative insights highlight four main concerns, limited capacity under delivery pressure, missing and hard‑to‑access evidence, uncertainty in interpreting AI Act obligations, and a lack of systematic mapping and validation tools, alongside clear interest in agent‑based closed‑loop support under strict HITL and transparency constraints.
\end{framed}

%%%%%%%%%% SUBGROUP ANALYSIS %%%%%%%%%%%%
\subsection{Exploratory Subgroup Patterns}
Given the small sample size, we conducted exploratory subgroup analyses to identify qualitative patterns and contrasts between selected subgroups. We examined three subgroups: (i) respondents working in high-risk contexts versus others, (ii) requirements engineers versus other roles, and (iii) sectoral groups (research, consulting, industry). Following the approach described in Section~\ref{sec:method}, we compared these subgroups using descriptive statistics and cross-tabulations of key items as well as three composite 0–100 indices (AI‑Act readiness, compliance burden, and AI/agent acceptance), and interpret observed differences qualitatively rather than as statistically significant effects.

\paragraph{High-Risk versus Non-High-Risk Contexts}
Respondents working in high-risk contexts under the EU AI Act tend to rate the Act as more important and report more advanced internal discussions. They also emphasize legal and liability risks more strongly and express lower tolerance for autonomous agent decisions without human approval. In contrast, respondents outside high-risk settings focus more on efficiency gains and have a slightly higher tolerance for errors. On our 0--100 indices, perceived AI/agent acceptance is similar in both groups (High-Risk: 71, Non-High-Risk: 75), and both report a comparable compliance burden (50 in each group), whereas High-Risk respondents indicate substantially lower AI-Act readiness than Non-High-Risk respondents (40 vs.\ 60).

\paragraph{Requirements Engineers versus Other Roles}
We compared respondents who self-identify as requirements engineers with other roles. RE tend to prioritize traceability, citation fidelity, and audit logs, and prefers co-pilot modes with mandatory human approval. They are also particularly interested in delegating RE drafting, structuring of requirements, and generation of testable acceptance criteria to agents, provided that guardrails and review gates are in place. Other roles place slightly more emphasis on efficiency gains and integration with existing delivery processes. Quantitatively, requirements engineers report slightly higher AI/agent acceptance (77 vs.\ 71), substantially lower perceived compliance burden (30 vs.\ 50), and higher AI-Act readiness (60 vs.\ 38) than other roles. Across both groups, willingness to support a pilot for an AACL remains high, conditional on visible improvements in time-to-compliance and coverage of relevant obligations.

\paragraph{Sectoral Patterns}
Finally, when grouping by sector, we observed differences in data and evidence practices and perceived use cases for agent-based support. Research organizations tend to have more complex, curated evaluation datasets and higher requirements for experimental reproducibility. Consultants emphasize the need for flexible, lightweight tools to accommodate heterogeneous customer environments. Industry and manufacturing contexts highlight technical documentation, traceability, and integration with existing quality management systems (QMS) and RE tool chains as key concerns. On the 0--100 indices, research organizations show the highest AI/agent acceptance (84 vs.\ 75 in industry and 70 in consulting), the highest perceived compliance burden (53 vs.\ 50 and 40), and the highest AI-Act readiness (60 vs.\ 50 and 37). Across all sectors, there is agreement that additional validation effort is expected and well-governed evidence bases are currently lacking.

\section{Discussion}
\label{sec:discus}
% Interpret the findings, linking them back to the original research questions.
% Compare results with existing literature.
% Address the implications of the findings and acknowledge limitations (e.g., sampling bias).
In this section, we discuss our findings in relation to two research questions: organizational preparedness for EU AI Act obligations in RE \textit{(RQ1)} and industry practitioners' perceptions of LLM-based validation tools \textit{(RQ2)}.

\subsection{Integrated Interpretation of Findings}
This subsection synthesizes the quantitative results, qualitative survey responses, and interview insights to provide an integrated picture of the findings. We address the three key aspects of the survey instrument: (i) organizational AI Act readiness in RE, (ii) current effort and perceived risks around mapping and validating regulatory obligations, and (iii) attitudes towards LLM-based, agentic closed-loop validation tools for requirements.

\paragraph{Aspect 1 (RQ1): Organizational AI Act Readiness}
Regarding RQ1, respondents are aware of the EU AI Act and generally consider it important, but organizational maturity is still low. Self‑reported knowledge is moderate, preparation is mostly limited to monitoring and early implementation efforts, and project‑level handling remains largely manual (e.g., documented reviews, controlled AI use). Governance structures and policies are only beginning to emerge. Qualitative insights underline a tension between recognizing the Act’s importance and lacking capacity for systematic implementation. High‑risk contexts stress governance and clear ownership, while smaller or non‑high‑risk organizations often address AI Act topics selectively or on a case-by-case basis. This pattern is consistent with our quantitative results in Figures~\ref{fig:ai-act-knowledge} and \ref{fig:ai-act-readiness-overview} and with prior observations on compliance readiness in regulated domains \cite{peraldifrati2010ReInSafetySys,kelly2024navigatingEuAiAct}.

\paragraph{Aspect 2 (RQ1): Current Effort and Perceived Risks}
For RQ1, there is a clear mismatch between expected validation effort and the robustness of available evidence. Many participants currently spend little time on mapping and validating regulatory obligations (almost half report \textless10 hours per month or none), yet those who forecast future effort expect a marked increase once the AI Act is fully applicable. The main perceived risks of non‑compliance are legal and liability exposure, reputational damage, and contractual or economic consequences. At the same time, building suitable data and evidence is hindered by high cost and effort, limited annotation quality and subgroup coverage, and restricted access to realistic customer data due to consent and contractual constraints. Lead times for evaluation datasets are often long or unknown, and synthetic data or simulations are only used under strict conditions. These challenges appear across sectors and roles and suggest that rising validation demands will hit fragile, costly, and slow‑to‑provision evidence bases. This is reflected in Figure~\ref{fig:risks-data-barriers}, which highlights legal and reputational risks and data barriers as dominant concerns.

\paragraph{Aspect 3 (RQ2): Attitudes towards LLM-based, Agentic Closed-Loop Validation Tools}
Regarding RQ2, practitioners are cautiously open to LLM‑based, agentic closed‑loop tools in RE. They expect quality and efficiency gains, particularly for drafting and structuring requirements, generating test cases, and improving documentation and traceability, and many plan to adopt such tools. However, they insist on transparency, security, and human control: accurate citations, human oversight, clear explanations, and safeguards such as access control and audit logging are seen as essential. The preferred mode is a collaborative co‑pilot, where humans retain final authority. Practitioners also expect support for identifying relevant law changes and extracting context‑specific legal requirements, but only under strong HITL and governance constraints, which directly inform the minimum requirements for an AACL validation agent. These attitudes are consistent with the trust, guardrail, and acceptance patterns shown in Figures~\ref{fig:ai-agent-guardrails} and \ref{fig:acceptance}, and they echo recent discussions on LLM‑based RE support \cite{Lai2025PreprintMAS4RE,jin2024mare,Hassine24LLMbasedTrace}.

\subsection{Minimum Requirements for an AACL Validation Agent}
Based on our findings, we have identified four minimum requirements that an AI Act-ready, agent-based closed-loop validation system must meet to be acceptable in practice.

\paragraph{Transparency and End-to-End Auditability}
To build trust, an AACL validation agent must provide end-to-end transparency and auditability. This means that every mapping between AI Act provisions and RE artifacts should be accompanied by high-fidelity source citations, and all prompts, model outputs, human feedback, and approvals must be captured in an immutable, queryable audit trail.

\paragraph{Guardrails and Security by Design}
Given the sensitivity of requirements and regulatory evidence, an AACL validation agent must have robust, built-in guardrails. This includes access control based on least privilege and roles, on-premise or regionally constrained processing, and an immutable audit log. The agent should also enforce data minimization and protection of personally identifiable information (PII) by design.

\paragraph{HITL Oversight and Constrained Autonomy}
The preferred operational mode for an AACL validation agent is a co-pilot with mandatory human approval. The agent should operate under strict security and data protection constraints, enforce fine-grained role and access management, and implement data minimization and PII protection by design. The agent should also implement event-based HITL triggers, which require explicit human review and approval when risk, uncertainty, impact, or missing evidence exceed defined thresholds.

\paragraph{Ability to Cope with Limited and Imperfect Data}
An AACL validation agent must be able to operate under partial and imperfect evidence. The agent should explicitly represent uncertainty, flag coverage gaps and evidence weaknesses, and make conservative recommendations when data are weak or missing. The agent should also support the incremental accumulation and reuse of curated evaluation snippets and evidence over time, without assuming the existence of large, fully representative gold-standard datasets at the outset.

In summary, our findings suggest that organizations are at an early stage of EU AI Act readiness and view LLM-based, agentic tools as promising but acceptable only under strong transparency, security, and human-oversight constraints. These insights inform a first set of minimum requirements for an AACL validation agent, which we will outline further in the conclusion.
\section{Conclusion}
\label{sec:conclu}
%This section summarizes our main results, answers the research questions, and outlines implications and directions for future work.

We investigated the hypothesis whether an LLM‑based, agentic validation tool can support experts in mapping EU AI Act requirements to project requirements and generating auditable, traceable RE artifacts. To explore this hypothesis, we used a mixed‑methods design with 10 expert interviews and an online survey (N=15) to assess organizational readiness, compliance challenges, and acceptance of such tools

For \textbf{RQ1} (organizational preparedness), our findings show that while practitioners are aware of the Act and consider it important, overall readiness remains limited. Most companies are at an early stage of monitoring and implementation, with few having fully implemented and continuously monitored policies. Participants expect compliance effort to increase substantially as the Act becomes fully applicable, yet struggle with limited, costly, and hard-to-access data and evidence.

For \textbf{RQ2} (perceptions of LLM-based, agentic closed-loop validation tools), we observe cautious but notable openness. Many respondents are interested in AI-supported tools for drafting and structuring requirements, generating test cases, and improving documentation and traceability. However, this interest is conditional on strong safeguards, including transparency of sources, end-to-end auditability, strict access control and data protection, and HITL oversight; fully autonomous operation is largely rejected outside low-risk settings.

We identified minimum requirements for LLM‑supported, agent‑based validation tools: transparency and auditability with clear sources, strong security and data protection, and HITL oversight.  Future research should develop and evaluate proof-of-concept systems in real-world projects to refine these requirements and assess their effectiveness and adoption.

\section*{Acknowledgment}
We gratefully acknowledge the Department of AI Assurance and Assessments (AAA) and the Lamarr Innovation Group for their collaborative effort, generous funding, and expert guidance, which were indispensable to the successful realization of Project AI Gov for RE.

We thank all interview participants for their time, insight, and willingness to contribute to our study, as well as the anonymous reviewers whose thorough feedback and constructive suggestions significantly enhanced the quality of this paper.

We acknowledge the use of AI model GPT-5.1 for editing and grammar enhancement purposes in this manuscript. While the AI model assisted with refining the language and structure, the intellectual content and ideas presented in this paper remain the sole responsibility of the authors.

\newpage
% \bibitem[XXX]{XXX}
% \end{thebibliography}
\bibliographystyle{ieeetr}
\bibliography{references}

\end{document}